\documentclass[prd,
,superscriptaddress,
nofootinbib,%
tightenlines ]{revtex4}
\usepackage{epsfig}
\usepackage[colorlinks=true,linkcolor=blue,urlcolor=blue,citecolor=blue]{hyperref}
\usepackage{amsfonts}
\usepackage{amssymb}

\newcommand{\ben}{\begin{displaymath}}
\newcommand{\een}{\end{displaymath}}
\newcommand{\be}{\begin{equation}}
\newcommand{\ee}{\end{equation}}
\newcommand{\bea}{\begin{eqnarray}}
\newcommand{\eea}{\end{eqnarray}}

\begin{document}
\title{Chiral theory of $\rho$-meson gravitational form factors}
\author{E.~Epelbaum}
 \affiliation{Institut f\"ur Theoretische Physik II, Ruhr-Universit\"at Bochum,  D-44780 Bochum,
 Germany}
\author{J.~Gegelia}
 \affiliation{Institut f\"ur Theoretische Physik II, Ruhr-Universit\"at Bochum,  D-44780 Bochum,
 Germany}
 \affiliation{High Energy Physics Institute, Tbilisi State
University, 0186 Tbilisi, Georgia}
\author{U.-G.~Mei\ss ner}
 \affiliation{Helmholtz Institut f\"ur Strahlen- und Kernphysik and Bethe
   Center for Theoretical Physics, Universit\"at Bonn, D-53115 Bonn, Germany}
 \affiliation{Institute for Advanced Simulation, Institut f\"ur Kernphysik
   and J\"ulich Center for Hadron Physics, Forschungszentrum J\"ulich, D-52425 J\"ulich,
Germany}
\affiliation{Tbilisi State  University,  0186 Tbilisi,
 Georgia}
 \author{M. V.~Polyakov}
  \affiliation{Institut f\"ur Theoretische Physik II, Ruhr-Universit\"at Bochum,  D-44780 Bochum,
 Germany}
\affiliation{Petersburg Nuclear Physics Institute, 
		Gatchina, 188300, St.~Petersburg, Russia}

\date{September 22, 2021}
\begin{abstract}
The low-energy chiral effective field theory of vector mesons
and Goldstone bosons in external gravitational field is considered.
The energy-momentum tensor is obtained and the gravitational form factors of the $\rho$-meson
are calculated up to next-to-leading order in the chiral expansion. 
This amounts to considering tree-level and one-loop order diagrams.
The chiral expansion of the form factors at zero momentum transfer as well as of the
slope parameters is worked out. Also, the long-range behaviour of the energy and internal
force distributions  is obtained and analysed. 
\end{abstract}

\maketitle

\bigskip
\bigskip
\begin{center}
{\large During the preparation of this paper Maxim Polyakov passed
  away. \\[5pt] We dedicate this paper to the memory of Maxim.} 
\end{center}

\bigskip
\bigskip

\section{Introduction}

The linear response of a hadron to a  change of the  background space-time metric is described
by the gravitational form factors (GFFs). For the first time, the  GFFs for spin-0  and spin-1/2
particles were introduced and discussed in detail
in Refs.~\cite{Kobzarev:1962wt,Pagels:1966zza}, for spin-1 particles
in Ref.~\cite{Holstein:2006ud}, and for hadrons  with arbitrary spin  in the recent work of
Ref.~\cite{Cotogno:2019vjb}.
The GFFs contain rich information about the internal structure of hadrons, such as the distribution
of the spin \cite{Ji:1996ek},  the energy distribution  
\cite{Ji:1994av}, as well as the  elastic pressure and shear force distributions
\cite{Polyakov:2002yz}. For recent reviews see Refs.~\cite{Polyakov:2018zvc,Lorce:2018egm}.

Our aim here is to study the GFFs of the spin-1 $\rho$-meson in  chiral effective field theory (EFT).
Hadrons with spin $S> 1/2$ are not spherically symmetric. The
spin, energy, and force distributions acquire higher multipole components (quadrupole, etc.)
\cite{Cotogno:2019vjb,Polyakov:2018rew,Polyakov:2019lbq,Panteleeva:2020ejw}. The higher
multipole energy and force distributions
carry valuable information about the mechanisms of the  hadron's binding. For example,
the large-$N_c$  picture  of baryons as chiral solitons implies certain relations between
the quadrupole energy and the force distributions \cite{Panteleeva:2020ejw,Kim:2020lrs}.
Experimental checks of these relations would allow one to reveal the nature of higher spin baryons.  

The GFFs of the $\rho$-meson were computed in the light-cone constituent quark model \cite{Sun:2020wfo}.
More recently, the gluon part of the GFFs was obtained in lattice  QCD calculations
\cite{Pefkou:2021fni}. Here, we investigate the dependence of the $\rho$-meson GFFs on the soft
scales (pion mass, small momentum transfer) using  chiral EFT.
To this end, following the logic of Ref.~\cite{Alharazin:2020yjv}, we first write down the
chiral effective action for the $\rho$- and $\omega$-mesons and pions in an
external gravitational field.  Next we obtain the corresponding energy-momentum tensor (EMT)
and compute the chiral corrections to the GFFs of the $\rho$-meson.
The corresponding calculation, in particular, allows us to obtain the large distance
behaviour of the energy and force distributions. 
The results of our study can be also used in chiral extrapolations of the lattice-QCD simulations
down to the physical values of  the pion masses.

Chiral EFTs with heavy degrees of freedom encounter a 
non-trivial power-counting problem \cite{Gasser:1987rb}.
In the one-nucleon sector of baryon chiral perturbation theory this problem can be
solved by applying the heavy-baryon approach
\cite{Jenkins:1990jv,Bernard:1992qa} or a suitably chosen renormalization condition
\cite{Tang:1996ca,Becher:1999he,Gegelia:1999gf,Fuchs:2003qc}.
   Because of the small nucleon-delta mass difference, the
$\Delta$ resonance can also be consistently included in the framework of EFT
\cite{Hemmert:1997ye,Pascalutsa:2002pi,Bernard:2003xf,Hacker:2005fh}. 

   The treatment of the $\rho$ meson in chiral EFT is 
complicated as it decays in two pions with masses that vanish in the chiral limit.
Because of this, for energies of the order of the $\rho$-meson mass, loop diagrams
develop large imaginary parts.  In distinction to  the baryonic sector,
these large power-counting-violating contributions
cannot be absorbed in the redefinition of the parameters of the Lagrangian as long as the
usual renormalization procedure is used. Still, the problem can be handled \cite{Djukanovic:2009zn}
by using the complex-mass renormalization scheme 
\cite{Stuart:1990,Denner:1999gp},
which is an extension of the on-mass-shell renormalization scheme
to unstable particles. For more details on and different approaches to these problems,
see e.g. Refs.~\cite{Ecker:1989yg,Borasoy:1995ds,Bruns:2004tj,Bruns:2008,Jenkins:1995vb,Bijnens:1996kg,Bijnens:1997ni,Bijnens:1997rv,Bijnens:1998di}.

Our work is organized as follows. In Section~\ref{Lagrdefs} we write down the action
corresponding to the effective Lagrangian up to next-to-leading order and obtain
the pertinent EMT. 
In Section~\ref{RaPC} we briefly discuss the renormalization and the power counting.
The definition and the calculations of the GFFs of the $\rho$ meson are presented in Section~\ref{GFFs}.
It also contains various chiral expansions of the 
obtained results. Section~\ref{EaFD} is devoted to the discussion of  the energy and the
force distributions, and  we summarize the obtained results in Section~\ref{summary}.

\section{Effective Lagrangian and the energy-momentum tensor}
\label{Lagrdefs}

Using the results of Refs.~\cite{Donoghue:1991qv,Djukanovic:2009zn}  we consider the following
action of $\rho$ and $\omega$ mesons and pions using the parametrization of the model~III
of Ref.~\cite{Ecker:1989yg} (where the $\rho$-meson vector fields
transform inhomogeneously under chiral transformations), interacting with an
external gravitational field  $g^{\mu\nu}$:
\begin{eqnarray}
S & = & \int d^4x \sqrt{-\text{\cal{g}}}\, \Biggl\{ \frac {F^2}{4}\, g^{\mu\nu}\, {\rm Tr} ( D_\mu U  (D_\nu U)^\dagger ) + \frac{F^2}{4}\,{\rm Tr}(\chi U^\dagger +U \chi^\dagger) \nonumber\\
&-& \frac{1}{2}\, g^{\mu\nu} g^{\alpha\beta}{\rm
Tr}\left(\rho_{\mu\alpha}\rho_{\nu\beta}\right) + g^{\mu\nu} \left[ M_{R}^2 +
\frac{c_{x} {\rm Tr} \left( \chi 
U^\dagger+U \chi^\dagger\right) }{4}\right]
{\rm Tr}\left[\left(
\rho_\mu-\frac{i\,\Gamma_\mu}{g}\right)\left(
\rho_{\nu}-\frac{i\,\Gamma_\nu}{g} \right)\right] \nonumber\\
&-& \frac{1}{4} g^{\mu\nu} g^{\alpha\beta} \left(
\partial_\mu\omega_{\alpha}-\partial_\alpha\omega_{\mu}\right)\left(
\partial_\nu\omega_\beta-\partial_\beta\omega_\nu\right)+ g^{\mu\nu} \frac{M_{\omega}^2\,
\omega_{\mu}\omega_\nu}{2} 
+ \frac{g_{\omega\rho\pi}}{2 \sqrt{-\text{\cal{g}}}}\,
\epsilon^{\mu\nu\alpha\beta}\, \omega_{\nu}\,
{\rm Tr}\left(\rho_{\alpha\beta} u_\mu
\right) \nonumber\\
&+& \left[v_{1}+v_{2} {\rm Tr} \left( \chi 
U^\dagger+U \chi^\dagger\right) \right] R \, {\rm Tr} (\rho_\mu \rho^\mu )   +v_{3} R^{\mu\nu} \, {\rm Tr} (\rho_\mu \rho_\nu ) \nonumber\\
&+& v_4 R \, {\rm
Tr}\left(\rho_{\alpha\beta}\rho^{\alpha\beta}\right) + v_5 R^{\mu\nu} \, g^{\alpha\beta}{\rm
Tr}\left(\rho_{\mu\alpha}\rho_{\nu\beta}\right)  + v_6 R^{\mu\nu\alpha\beta} \,  {\rm
Tr}\left(\rho_{\mu\nu}\rho_{\alpha\beta}\right) 
 \Biggr\},
\label{PionAction}
\end{eqnarray}
where
\begin{eqnarray}
U&=&u^2={\rm exp}\left(\frac{i\vec{\tau}\cdot\vec{\pi}}{F}\right),
\nonumber\\
\rho^\mu & = & \frac{\vec\tau\cdot\vec\rho\,^\mu}{2},\nonumber\\
\rho^{\mu\nu} & = &
\partial^\mu\rho^\nu-\partial^\nu\rho^\mu - i
g \left[\rho^\mu,\rho^\nu\right] \,,\nonumber\\
\chi & = & 2 B_0(s+i p) \,,\nonumber\\ 
D_\mu U & = & \partial_\mu U -i r_\mu U +i U l_\mu\,,\nonumber\\
\Gamma_\mu &= & \frac{1}{2}\,\bigl[ u^\dagger\partial_\mu u+u
\partial_\mu u^\dagger
- i\,\left( u^\dagger r_\mu u +u l_\mu u^\dagger\right)
\bigr]\,, \nonumber
\\
u_\mu & = & i \left[ u^\dagger \partial_\mu u-u \partial_\mu
u^\dagger -i(u^\dagger r_\mu u -u l_\mu u^\dagger ) \right]. \label{somedefinitions}
\end{eqnarray}
Terms involving the Riemann-tensor $R^{\mu \nu \alpha \beta}$, the
Ricci-tensor $R^{\mu \nu}$  and the Ricci scalar $R$ (for definitions of these quantities see, e.g, Ref.~\cite{Birrell:1982ix}) are those
with non-minimal coupling of the $\rho$-meson fields to gravity, which are relevant for the considered order of accuracy.
The parameter $B_0$ is proportional to the 
scalar vacuum condensate and $s$, $p$, $l_\mu =v_\mu-a_\mu $ and $r_\mu =v_\mu + a_\mu $ 
are external sources, while $F$ denotes the pion-decay constant in the chiral
limit. Further, $M_R^2$ and $M_\omega^2$ are the (complex) pole positions of $\rho$ and $\omega$ propagators in chiral limit,
and the $v_i \ (i=1,\ldots,6),$ $g$, $c_x$, and $g_{\omega\rho\pi}$ are coupling constants.
For the $\rho\pi\pi$ coupling we use \cite{Meissner:1987ge} 
\begin{eqnarray}
M_R^2 & = & a\, g^2 F^2 \,, \label{M0}
\end{eqnarray}
which in the case of $a=2$ amounts to the KSFR relation
\cite{Kawarabayashi:1966kd,Riazuddin:sw,Djukanovic:2004mm}.
Although phenomenologically $a \simeq 2$, in what follows we will keep this paramter explicitely. 
All parameters of the effective Lagrangian are to be interpreted as renormalized ones. 
We apply the complex-mass scheme and do not show counterterms explicitly, however,
their contributions are taken into account in calculations of the quantum corrections to
the physical quantities. 

\medskip

Applying the standard formula for the EMT of matter fields interacting with the metric fields \cite{Birrell:1982ix}
\begin{eqnarray}
T_{\mu\nu} & = & \frac{2}{\sqrt{-g}}\frac{\delta S }{\delta g^{\mu\nu}}\,,
\label{EMTMatter}
\end{eqnarray}
to the action of Eq.~(\ref{PionAction}) we obtain  in flat spacetime:
\begin{eqnarray}
T_{\mu\nu} & = &  \frac {F^2}{4}\, {\rm Tr} ( D_\mu U  (D_\nu U)^\dagger +D_\nu U  (D_\mu U)^\dagger ) \nonumber\\
&-& \eta_{\mu\nu} \left\{ \frac {F^2}{4}\, {\rm Tr} ( D^\alpha U  (D_\alpha U)^\dagger ) +  \frac{F^2}{4}\,{\rm Tr}(\chi U^\dagger +U \chi^\dagger) \right\} \nonumber\\
&-& 2\, \eta^{\alpha\beta}{\rm
Tr}\left(\rho_{\mu\alpha}\rho_{\nu\beta}\right) + 2 \left[ M_{\rho}^2 +
\frac{c_{x}{\rm Tr} \left( \chi 
U^\dagger+U \chi^\dagger\right) }{4}\right]
{\rm Tr}\left[\left(
\rho_\mu-\frac{i\,\Gamma_\mu}{g}\right)\left(
\rho_{\nu}-\frac{i\,\Gamma_\nu}{g} \right)\right] \nonumber\\
&-& \eta_{\mu\nu} \left\{ - \frac{1}{2}\, {\rm
Tr}\left(\rho_{\alpha\beta}\rho^{\alpha\beta}\right) + \left[ M_{\rho}^2 +
\frac{c_{x}{\rm Tr} \left( \chi 
U^\dagger+U \chi^\dagger\right) }{4}\right]
{\rm Tr}\left[\left(
\rho_\alpha-\frac{i\,\Gamma_\alpha}{g}\right)\left(
\rho^{\alpha}-\frac{i\,\Gamma^\alpha}{g} \right)\right]  \right\} \nonumber\\
&-& \eta^{\alpha\beta} \left(
\partial_\mu\omega_{\alpha}-\partial_\alpha\omega_{\mu}\right)\left(
\partial_\nu\omega_\beta-\partial_\beta\omega_\nu\right)+  M_{\omega}^2\,
\omega_{\mu}\omega_\nu ,\nonumber\\
&-& \eta_{\mu\nu} \left\{  - \frac{1}{4}  \left(
\partial_\alpha\omega_{\beta}-\partial_\beta\omega_{\alpha}\right)\left(
\partial^\alpha\omega^\beta-\partial^\beta\omega\alpha\right)+ \frac{M_{\omega}^2\,
\omega_{\alpha}\omega^\alpha}{2} \right\} \nonumber\\
&+& 2(\eta_{\mu\nu} \partial^2-\partial_\mu\partial_\nu) \left\{ \left[v_{1}+v_{2} {\rm Tr} \left( \chi 
U^\dagger+U \chi^\dagger\right) \right] {\rm Tr} (\rho_\alpha \rho^\alpha )  +v_4 {\rm
Tr}\left(\rho_{\alpha\beta}\rho^{\alpha\beta}\right) \right\} \nonumber\\
&+&(\eta_{\mu\alpha} \eta_{\nu\beta} \partial^2 + \eta_{\mu\nu}\partial_\alpha\partial_\beta- \eta_{\mu\alpha}\partial_\nu\partial_\beta- \eta_{\nu\alpha}\partial_\mu\partial_\beta)\left[ v_{3}  {\rm Tr} (\rho^\alpha \rho^\beta ) 
+  v_{5} \, \eta_{\lambda\sigma} {\rm Tr} (\rho^{\alpha\lambda}\rho^{\beta\sigma})\right]  \nonumber\\
&+& 4 \, v_6 \, \eta^{\alpha \lambda }
    \eta^{\beta \sigma} \partial_ {\lambda }  \partial_ {\sigma }  \,  {\rm Tr} (\rho_{\mu\beta}\rho_{\nu\beta})
\,,
\label{PionEMT}
\end{eqnarray}
where $\eta_{\mu\nu}$ is the metric tensor in  Minkowski space.

\section{Renormalization and Power Counting}
\label{RaPC}
To perform the renormalization we express the bare quantities in terms of
renormalized ones and  counterterms  and  apply the complex-mass renormalization scheme
\cite{Stuart:1990,Denner:1999gp}.
We parameterize the pole of the $\rho$-meson dressed propagator in the chiral limit
as $M_R^2=(M_\chi - i\, \Gamma_\chi/2)^2$, where $M_\chi$ and
$\Gamma_\chi$ are the pole mass and width of the $\rho$ meson
in the chiral limit, respectively.  Both are input parameters within our formalism.

Following Ref.~\cite{Djukanovic:2009zn}, we fix the mass counterterm and the wave function renormalization
constant  by requiring
that in the chiral limit, $M_R^2$ coinsides with the pole position of the dressed propagator and the residue 
is equal to unity.  The renormalized complex mass $M_R$ appears in the propagator, and the
counterterms are included perturbatively.  Notice that in the complex-mass renormalization scheme,
the counterterms are also complex quantities. This does, however, not lead to a violation of
unitarity as one might naively expect \cite{Bauer:2012gn,Denner:2014zga}.
Let us demonstrate this using the example of the renormalization of the $\rho$-meson mass.
The Lagrangian is given in terms of bare parameters, and 
physical quantities also can be calculated in terms of these parameters within some
ultraviolet regularization scheme. The physical mass of a stable particle as well as the mass
and width of an unstable particle can be obtained from the 
corresponding two-point function by finding its pole position. 
Defining the self-energy of the $\rho$-meson as the sum of all one-particle irredusible diagrams
contributing  to the two-point function of the $\rho$-meson field operators we parameterize this
quantity as 
\begin{equation}
i \Pi^{\mu\nu}(p) = i \left( g^{\mu\nu} \Pi_1(p^2) +  p^\mu p^\nu \Pi_2(p^2)\right)\,. 
\label{seeq}
\end{equation}
The equation determining the pole position $z$ of the two-point function written in terms of the
bare parameters has the form:
\begin{equation}
z-M_0^2 -\Pi_1(z, M_0, M_\pi,\cdots ) =0\,,
\label{poleeq}
\end{equation}
where $M_0$ is the bare mass of the $\rho$-meson, $M_\pi$ is the pion  mass, and the ellipses denote
other parameters of the Lagrangian and also the ultraviolet regulator.
The solution to Eq.~(\ref{poleeq})   has the form:
\begin{equation}
z= f (M_0, M_\pi ,  \cdots) \equiv M_0^2 +{\rm corrections }\,.
\label{solpole}
\end{equation}
We denote the quantity $z$ in chiral limit by $M_R^2$ and invert Eq.~(\ref{solpole}) for $M_\pi =0$ 
to obtain
 \begin{equation}
M_0^2 = f^{-1} (M_R^2 , 0,  \cdots)  \equiv M_R^2 + \hbar \, \delta M^2 (\hbar, M_R^2,\cdots ) \,,
\label{bareM}
\end{equation}
where $M_R^2$ and  $\delta M^2 $ are both complex, while $M_0^2$ is real. We have indicated  the
explicit factor of $\hbar $ to emphasize that within the formalism
employed in the current work, 
after substituting Eq.~(\ref{bareM}) in the Lagrangian, we treat the second term  on the right-hand 
side (after further expanding it in powers of $\hbar$) together with the loop diagrams, i.e.
perturbatively.

\medskip

For the effective Lagrangian, we apply the standard rules counting the pion mass and the
derivatives acting on pion fields as small quantities, while the derivatives acting on the
heavy vector mesons count as large quantities of order ${\cal O}(1)$.
The large mass of the $\rho$-meson that does not vanish in chiral limit  
violates the simple correspondence between the power counting for the Lagrangian and the
power counting for the loop diagrams, thus leading to a considerable complication. 
One needs to investigate all possible flows of the external momenta through the
internal lines of the considered loop diagram.  Next, assigning powers to propagators and vertices,
one needs to determine the chiral order for a given flow of external momenta.
Finally, the smallest order resulting from all possible assignments
should be defined as the chiral order of the given diagram \cite{Djukanovic:2009zn}.
To assign the corresponding chiral order to a diagram for a given flow of external momenta we
apply the following rules: Taking $q$ as a small quantity like the pion mass or small external momenta, 
pion propagators count as ${\cal O}(q^{-2})$ if  not carrying large
external momenta while ${\cal O}(q^{0})$
otherwise. A vector-meson propagator counts as ${\cal O}(q^{0})$
if it does not carry large external momenta and as ${\cal O}(q^{-1})$ if it does.
The vector-meson mass counts as ${\cal O}(q^{0})$, the width of the vector mesons as well as the
pion mass count as ${\cal O}(q^{1})$.
Interaction vertices generated by the effective Lagrangian of the order $n$ do not automatically count
as ${\cal O}(q^n)$ but rather need to be assigned orders according to
a given flow of large and
small external momenta. As the contributions of loops involving only vector meson propagators can
be absorbed systematically in the redefinition of the parameters of the effective Lagrangian,
such loop diagrams need not be included at low energies.

\section{Gravitational form factors  of the $\rho$ meson: Definitions and calculation}
\label{GFFs}

The gravitational form factors (GFFs) of a spin-1 particle were defined for the first time
in Ref.~\cite{Holstein:2006ud}. Here, we follow the conventions
and notations of Ref.~\cite{Polyakov:2019lbq}, in which the GFFs of a spin-1 particle were defined as:
\begin{eqnarray}
\langle p', \sigma'| T_{\mu\nu}| p,\sigma \rangle &=& \epsilon^{* \alpha'} (p',\sigma') \epsilon^\alpha (p,\sigma) 
\Biggl[ 2 P_\mu P_\nu  \left( -\eta_{\alpha\alpha'} A_0(t)  + \frac{P_\alpha P_{\alpha'}} {m^2}\,A_1(t)   \right) \nonumber\\
&& + 2 \left[ P_\mu ( \eta_{\nu \alpha'} P_{\alpha}+\eta_{\nu\alpha} P_{\alpha'}) + P_\nu (\eta_{\mu \alpha'} P_{\alpha}+\eta_{\mu\alpha} P_{\alpha'} )\right] J(t)  \nonumber\\
&+&  \frac{1}{2} (\Delta_\mu \Delta_\nu-\eta_{\mu\nu} \Delta^2) \left( \eta_{\alpha\alpha'}D_0(t) + \frac{P_\alpha P_{\alpha'}} {m^2}\, D_1(t) \right) \nonumber\\
&+& \biggl[ \frac{1}{2} \, (\eta_{\mu \alpha}  \eta_{\nu \alpha'} + \eta_{\mu \alpha'}  \eta_{\nu \alpha}  )\Delta^2 
- ( \eta_{\nu \alpha'} \Delta_\mu   + \eta_{\mu \alpha'}  \Delta_\nu  ) P_\alpha  \nonumber\\
&+& ( \eta_{\nu \alpha} \Delta_\mu   + \eta_{\mu \alpha}  \Delta_\nu  ) P_{\alpha'} -4 \, \eta_{\mu\nu} P_\alpha P_{\alpha'}
\biggr] E(t) \Biggr] \,,
\label{RGFFsdef}
\end{eqnarray}
where $\Delta=p_f - p_i$, $P=(p_f + p_i)/2$, $m$ is the mass (note that we reserve the symbol $M$ for
the $\rho$ and $\omega$ masses) of the spin-1 particle and the polarization
vector $\epsilon_\alpha(p,\sigma)$ satisfies the condition
\begin{equation}
\sum_\sigma \epsilon_\alpha(p,\sigma) \epsilon_\beta(p,\sigma) =-\eta_{\alpha\beta}+\frac{p^\alpha p^\beta}{m^2}\,.
\label{sumofpols}
\end{equation} 
For the reader's convenience we collect in Table~\ref{emtffs} other notations  for the GFFs
of spin-1 particles used in the literature.
\begin{table*}
\caption{\label{emtffs} The notations for GFFs of spin-1 particles used in the literature.  
}
\begin{center}
	\begin{tabular}{c|cccccc}
	\hline
	\hline \\
	$\text{\cite{Polyakov:2019lbq} and this work}$ & $A_0$ & $A_1$ & $D_0$ & $D_1$ & $J$ & $E$  \\ \hline \\
	$\text{Holstein~\cite{Holstein:2006ud}}$ & $F_1$ & $4F_5$ & $-2F_2$ & $8F_6$ & $F_3$ & $-2F_4$ \\ \hline \\
	$\text{Abidin et al.~\cite{Abidin:2008ku}}$ & $A$ & $-2E$ & $C$ & $-8F$ & $A+B$ & $D$  \\ \hline \\
	$\text{Taneja et al.~\cite{Taneja:2011sy}}$ & ${\cal G}_{1}$ & $-2{\cal G}_{2}$ & $-{\cal G}_{3}$ & $-2{\cal G}_{4}$ & $\frac12{\cal G}_{5}$ & $-\frac12{\cal G}_{6}$ \\ \hline \\
	$\text{Cosyn et al.~\cite{Cosyn:2019aio}}$ & ${\cal G}_{1}$ & $-2{\cal G}_{2}$ & $-{\cal G}_{3}$ & $-2{\cal G}_{4}$ & $\frac12{\cal G}_{5}$ & $-\frac12{\cal G}_{6}$  \\ \hline \hline \\
	$\text{Cosyn et al.~\cite{Cosyn:2018thq} generalized form factors}$ & $A^a_{2,0}$ & $-2C^a_{2,0}$ & $-4F^a_2$ & $-8G^a_2$ & $\frac12 B^a_{2,0}$ & $D^a_{2,1}$  \\
	\hline
	\hline
	\end{tabular}
\end{center}
\end{table*}

As the $\rho$-meson is an unstable particle we extract its gravitational form factors from the
residue at the complex double-pole of the three-point correlation function of  the EMT 
and the vector meson fields \cite{Gegelia:2010nmt}.
In this case, $m^2$ in the above formulas is the complex pole position of the corresponding dressed propagator.

\begin{figure}
\epsfig{file=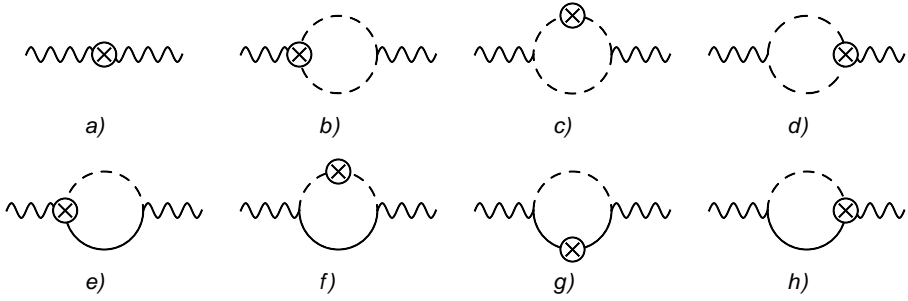, width=12truecm}
\caption[]{\label{RFF:fig} Tree-level and one-loop diagrams contributing to the
$\rho$-meson gravitational form factors. 
The dashed, solid,
and wiggly lines correspond to the pion, the $\omega$-meson, and
the $\rho$-meson, respectively. The crossed vertex denotes an
insertion of the EMT.
}
\end{figure}

\medskip

In the current work we consider contributions  of tree-level and one-loop diagrams
to the gravitational form factors of the $\rho$-meson, see Fig.~\ref{RFF:fig}.
At tree level, there are contributions of higher-order terms in the effective Lagrangian. As the
higher-order Lagrangian is not available yet, we include these
contributions to the form factors parametrized in the
general form as polynomials of  the pion mass and the momentum transfer squared.
To calculate the loop diagrams we apply dimensional regularization and use the program
FeynCalc \cite{Mertig:1990an,Shtabovenko:2016sxi}. 
To calculate the  various expansions of the loop integrals we applied the method of dimensional
counting of Ref.~\cite{Gegelia:1994zz}. 
To simplify the analytic expressions we take $M_\omega^2 = M_R^2$, which is a
good approximation given the accuracy of this work. 
Below, we specify the chiral expansions of the form factors at $t=0$ and of the slope parameters and
also provide expressions for the form factors in the small-$t$ region in the chiral limit. Within the accuracy
of our calculations, the pion mass at leading order in the chiral expansion  can be replaced by its
full expression $M_\pi$. 

\medskip

The chiral expansion of the form factors at zero momentum-transfer has
the form:
\begin{eqnarray}
A_0(0) & = & 1\,,\nonumber\\
A_1(0) & = & 8v_6 M_R^2  + X_{A_1} M_\pi^2  + \frac{ g_{\omega \rho \pi }^2 M_R}{48 \pi  F^2} \, M_\pi  -  \frac{ g_{\omega \rho \pi }^2 \left(1+ 4v_6 M_R^2 \right)  }{8
   \pi ^2 F^2}  \, M_\pi^2  \ln \frac{M_\pi}{M_R} + \mathcal{O} (M_\pi^3 )\,, \nonumber\\
J(0) & = & 1\,,\nonumber\\
D_0(0) & = & 1 +4v_1 + 8v_4 M_R^2 + X_{D_0} M_\pi^2 - \frac{ \left(a+ 3  g_{\omega \rho \pi }^2 (v_1+2v_4 M_\rho^2) \right) }{12 \pi ^2 F^2} \, M_\pi^2 \ln \frac{M_\pi}{M_R} + \mathcal{O} (M_\pi^3 ) \,, 
\nonumber\\
   D_1(0) & = & -8 \left(4v_4+v_5+v_6\right) M_R^2 
   +  \frac{ g_{\omega \rho \pi }^2 M_R^3}{60 \pi  F^2 } \,\frac{1}{M_\pi} +  \frac{M_R^2 \left(5
   g_{\omega \rho \pi }^2-8 \,a\right)  }{60 \pi ^2 F^2} \,  \ln \frac{M_\pi}{M_R}  + \mathcal{O} (M_\pi)
   \,, \nonumber\\ 
  E(0) & = &   1 -v_3 -v_5 M_R^2 + X_{E} M_\pi^2 + \frac{g_{\omega \rho \pi }^2 M_R}{96 \pi 
   F^2} \, M_\pi
    +  \frac{ \left( (6 \,a_3 +6\,v_5 M_R^2 -5) g_{\omega \rho \pi }^2-4\,a\right)   }{96 \pi ^2
   F^2} \, M_\pi^2  \ln
    \frac{M_\pi}{M_R}  + \mathcal{O} (M_\pi^3 )
    \,.
\label{D0}
\end{eqnarray} 
Here, $X_{F_i}$ (as well as $Y_{F_i}$, $Z_{F_i}$ and $W_{F_i}$ below) are some
linear combinations of renormalized complex-valued low-energy constants from the higher-order
effective Lagrangian.


The above equations provide the dependence of the GFFs at zero momentum transfer on the pion mass.
They can be used for extrapolations of the lattice-QCD results for the GFFs to
the physical values of the pion masses. In recent lattice calculations of the  gluon part of GFFs
for the $\rho$-meson \cite{Pefkou:2021fni}, it was found, in particular, that the value of
$D_1(0)$ is compatible with zero, albeit with large error
bars ($D_1(0)=0.0\pm 0.7$). From our calculations we see that $D_1(0)\sim 1/M_\pi$ for small pion  masses.
This singular contribution alone leads to a large
value of $D_1(0)\approx 4$ for the physical pion mass.  This value of the
singular part of the GFF can, unfortunately, not be
directly compared to the results of the lattice simulations of Ref.~\cite{Pefkou:2021fni} as only the
gluon part of the GFF $D_1^g$ was computed in that paper. However,
we expect that the singular $\sim 1/M_\pi$ part is also present in $D^g_1(0)$ with  a
slightly modified coefficient.
This suggests that the chiral extrapolation  of lattice results of Ref.~\cite{Pefkou:2021fni}
for the pion mass to its physical point should be studied with great care.

\medskip

Defining the slopes $s_{F}$ as the coefficients of linear terms in the Taylor expansion of the
form factors, $F(t)=F(0)+s_F \, t+\ldots$, 
we obtain for their chiral expansions the following results:
\begin{eqnarray}
s_{A_0} & = &  \frac{v_5}{2} +Z_{A_0} M_\pi^2 -\frac{ g_{\omega \rho \pi }^2}{64 \pi  F^2 M_R} \,M_\pi  -\frac{ g_{\omega \rho \pi }^2 \left(6v_5 M_R^2+7\right) }{192 \pi ^2 F^2 M_R^2} \, M_\pi^2 \ln \frac{M_\pi}{M_R}   + \mathcal{O} (M_\pi^3 )
   \,,\nonumber\\
s_{A_1} & = &  Y_{A_1} -\frac{g_{\omega \rho \pi }^2 M_R}{480 \pi  F^2} \,\frac{1}{M_\pi} + \mathcal{O} (M_\pi )
\,,\nonumber\\
s_J & = &  \frac{1}{2} \left(v_5+2v_6\right) +Z_{J} M_\pi^2   -\frac{ 17 g_{\omega \rho \pi }^2+a  }{1152 \pi ^2 F^2} 
    -\frac{g_{\omega \rho \pi }^2 }{192 \pi^2 F^2}\,\ln \frac{M_\pi}{M_R}  + \frac{5 g_{\omega \rho \pi  }^2}{768 \pi  F^2 M_R} \,M_\pi  \nonumber\\
&-& \frac{ g_{\omega \rho \pi }^2 \left(3v_5M_R^2+6v_6 M_R^2-5\right)-2\,a}{96 \pi ^2 F^2 M_R^2} \, M_\pi^2 \ln \frac{M_\pi}{M_R} + \mathcal{O} (M_\pi^3 )
   \,,
   \nonumber\\
s_{D_0} & = & -4v_4-\frac{v_5}{2} +Z_{D_0} M_\pi^2  +\frac{(35+24 i \pi ) a -36 g_{\omega \rho \pi }^2 }{1440 \pi ^2 F^2}   
+\frac{ g_{\omega \rho \pi }^2 M_R}{160 \pi  F^2}  \,\frac{1}{M_\pi}  
   + \frac{  5 g_{\omega \rho \pi }^2+4\,a  }{240 \pi ^2 F^2} \, \ln \frac{M_\pi}{M_R} \nonumber\\
 &+ & \frac{43 g_{\omega \rho \pi}^2}{3840 \pi  F^2 M_R}  \, M_\pi   +  
   \frac{ 5 g_{\omega \rho \pi }^2 \left(48v_4 M_R^2+6v_5 M_R^2+7\right)-8\,a  }{960 \pi ^2 F^2 M_R^2} \, M_\pi^2 \ln \frac{M_\pi}{M_R} + \mathcal{O} (M_\pi^3 ) \,,  \nonumber\\
  s_{D_1} & = &  Y_{D_1} +\frac{g_{\omega \rho \pi }^2 M_R^3}{560 \pi  F^2} \frac{1}{M_\pi^3} -\frac{\left(11 g_{\omega \rho \pi
   }^2-12\,a\right) M_R^2}{840 \pi ^2 F^2} \frac{1}{M_\pi^2}  -\frac{131 g_{\omega \rho \pi }^2 M_R}{13440 \pi  F^2} \frac{1}{M_\pi}  
 -  \frac{ 35 g_{\omega
   \rho \pi }^2+ 64 \,a}{840 \pi ^2 F^2} \,  \ln \frac{M_\pi}{M_R} + \mathcal{O} (M_\pi) \,,
                \nonumber\\
  s_E & = & \frac{1}{2} \left(v_5+2v_6\right) + Z_{E} M_\pi^2  +\frac{\left(35+12 i \pi \right) a -92 g_{\omega \rho \pi }^2}{5760 \pi ^2 F^2} -\frac{g_{\omega \rho \pi }^2 M_R}{1920 \pi  F^2} \frac{1}{M_\pi}  
  +\frac{  4a -5 g_{\omega \rho \pi }^2 }{960 \pi ^2 F^2} \,  \ln \frac{M_\pi}{M_R}
   \nonumber\\
  &+& \frac{19 \, g_{\omega \rho \pi }^2}{15360 \pi  F^2 M_R} \, M_\pi
  - \frac{ 5v_5 g_{\omega \rho \pi }^2 M_R^2+10v_6 g_{\omega \rho \pi }^2 M_R^2+2 \,a  }{160 \pi ^2 F^2 M_R^2} \, M_\pi^2 \ln \frac{M_\pi}{M_R}  + \mathcal{O} (M_\pi^3 )
  \,.
\label{Rs}
\end{eqnarray} 
We see from the above expressions that some of the slopes have strong singularities for  small
pion masses. This again underlines the need of
a careful analysis of the chiral extrapolation of lattice data. 


\medskip

It is also instructive to study the $t$-dependence of GFFs in the chiral limit $M_\pi=0$.
The corresponding results will allow us in the next 
section to derive the large distance asymptotics of the energy and force distributions. 
The expressions of the form factors in the small-$t$ region in the chiral limit has the form:
\begin{eqnarray}
A_0(t) &=&  1+\frac{v_5 }{2}\, t  + W_{A_0} t^2+\frac{7  g_{\omega \rho \pi }^2}{4096 F^2
   M_R} \, (-t)^{3/2} +\frac{5
   g_{\omega \rho \pi }^2 }{768 \pi ^2 F^2 M_R^2}  \, t^2  \ln \left(-\frac{t}{M_R^2}\right) 
   \,,
\nonumber\\
A_1(t) &=& 8 \,v_6 M_R^2 + Y_{A_1} t +  \frac{3 g_{\omega \rho \pi }^2 M_R}{1024 F^2} \, \sqrt{-t}  
\,,
\nonumber\\
 J(t) &=&  1+\frac{1}{2} \left(v_5+2v_6\right) t     -\frac{  9 g_{\omega \rho \pi }^2  +a }{1152 \pi ^2 F^2} \, t
 -\frac{ g_{\omega \rho \pi }^2  }{384 \pi ^2 F^2} \, t \ln
   \left(-\frac{t}{M_R^2}\right)
   \,,
\nonumber\\
D_0(t)&=&  1+4v_1 +8v_4 M_R^2-\left(4v_4+\frac{v_5}{2}\right) t
          -\frac{5  g_{\omega \rho \pi }^2 M_R}{1024 F^2} \, \sqrt{-t}
          + \frac{5 g_{\omega \rho \pi }^2+4\,a }{480 \pi ^2
   F^2} \,  t \ln \left(-\frac{t}{M_R^2}\right) \nonumber\\
&+& \frac{3 (7+40 i \pi ) a -200 g_{\omega \rho \pi }^2 }{7200 \pi ^2 F^2} \, t 
\,,
\nonumber\\
D_1(t) &=&  -8 \left(4v_4+v_5+v_6\right) M_R^2 +\frac{ M_R^2 \left(47 a-20 g_{\omega \rho \pi }^2\right)}{450 \pi ^2 F^2} 
 + \frac{3 g_{\omega \rho \pi }^2 M_R^3}{256 F^2}  \frac{1}{\sqrt{-t}} +  \frac{ \left(5 g_{\omega \rho \pi }^2 - 8\,a  \right) M_R^2}{120 \pi ^2 F^2}  \,  \ln \left(-\frac{t}{M_R^2} \right)  
   \,,
\\
E(t) &=&  1 -a_3 -v_5 M_R^2+\frac{1}{2} \left(v_5+ 2\,v_6\right) t  +\frac{ a(60 \pi i - 9) -245
   g_{\omega \rho \pi }^2 }{28800 \pi ^2 F^2} \,t  + \frac{g_{\omega \rho \pi }^2 M_R}{1024 F^2} \, \sqrt{-t} + \frac{ 4 a -5 g_{\omega
   \rho \pi }^2 }{1920 \pi ^2 F^2} \, t \ln
         \left(-\frac{t}{M_R^2}\right)\,.
         \nonumber
\label{ChexpFFs}
\end{eqnarray}  
Notice  that 
while not all analytic (at $t=0$)  terms can be absorbed into renormalization of
the coupling constants, all power counting violating pieces
are systematically removed. 


\section{Large distance behaviour of the energy and force distributions}
\label{EaFD}

It is particularly interesting to look at 
the energy distribution and mechanical properties such as the
elastic pressure and shear force distributions
inside the $\rho$-meson. These fundamental distributions are encoded
in the static
EMT defined in the Breit frame as \cite{Polyakov:2002yz}:
\be
\label{eq:BFEMT}
T^{\mu\nu} (\vec r, \sigma^\prime,\sigma ) 
= 
\int {d^3 \Delta \over  (2\pi)^3 \, 2E } \, e^{-i \vec \Delta \cdot \vec r} 
\langle p^\prime, \sigma^\prime \, |{\hat T}_{\rm QCD}^{\mu\nu}(0)|p,\sigma \rangle .
\ee
Here, ${\hat T}_{\rm QCD}^{\mu\nu}(0)$ is the QCD EMT operator of the matrix element that is
computed between hadron states with spin projections $\sigma, \sigma'$ and 
momenta $p^0=p^{0\prime}=E=\sqrt{m^2+\vec \Delta^2/4}$,  and $p^{i\prime}=-p^i=\Delta^i/2$. The
$00$-component of the static EMT contains the information
about the energy distribution, the  $0i$-components encode the spin distribution
while the $ik$-components provide us with  the distributions of elastic pressure and shear
forces inside the hadron  \cite{Polyakov:2002yz}. 

In relativistic quantum field theory it is impossible to localize
  an one-particle state with an
accuracy better than its Compton wave  length $\lambdabar= {\hbar}/{(m c)}$.
Therefore, at distances smaller or of the order of $\lambdabar$, one has to interpret the
Breit frame static  EMT of Eq.~(\ref{eq:BFEMT}) from a quasi-probabilistic phase-space
perspective \cite{Lorce:2018egm,Lorce:2020onh}.
The phase-space picture connecting the Breit frame to the light front has
been obtained, for the first time, in the study of the angular momentum distribution in Ref.~\cite{Lorce:2017wkb}.
The static EMT can be viewed as  the Wigner phase-space average \cite{Wigner:1932eb,Hillery:1983ms}
of the distributions of forces inside the nucleon in its rest-frame. See the detailed discussion
in Ref.~\cite{Lorce:2018egm} for the EMT and in Ref.~\cite{Lorce:2020onh} for the charge densities. Due
to Heisenberg's uncertainty principle, the Wigner distributions only have a 
quasi-probabilistic interpretation. For large distances $r\gg {1}/{(2m)}$ in Eq.~(\ref{eq:BFEMT}),
the Wigner distributions acquire a strict probabilistic interpretation, see the detailed discussion in
Refs.~\cite{Lorce:2018egm,Lorce:2020onh,Lorce:2017wkb}.

If one insists on a strict probabilistic
interpretation of distributions, the static EMT of Eq.~(\ref{eq:BFEMT}) for $r\sim 1/(2 m)$ acquires
the so-called relativistic corrections  discussed since the 1950ties~\cite{yennie}.
See Refs.~\cite{Burkardt:2000za,Miller:2010nz,Jaffe:2020ebz} for more recent discussion.
The relativistic  corrections can be kinematically suppressed if one considers the distributions
in the infinite momentum frame (IMF) or if one uses quantization on the light front, 
see, e.g., Refs.~\cite{Burkardt:2000za,Miller:2010nz}.
In the recent study \cite{Lorce:2020onh}, 
the natural interpolation between the Breit frame and IMF charge
distributions was obtained using the phase-space Wigner distributions.
Such an analysis can also be repeated for the case of the force distributions. 

The densities for the internal force distributions in the nucleon in
the IMF and on the light front were derived first in Ref.~\cite{Lorce:2018egm} 
with the help of the Wigner phase-space distribution. More recently, light-front force
distributions were also obtained in  Ref.~\cite{Freese:2021czn} using light-cone quantization methods. 
These densities possess a strict probabilistic interpretation (no relativistic corrections)
and they come out identical in both approaches. 

In this paper we are interested in the large-distance behaviour of the energy and the
force distributions ($r\gg {1}/{(2m)}$). 
The Breit-frame distributions  
do possess a probabilistic interpretation  in this limit. As the relativistic corrections are
parametrically suppressed in the range of applicability of the chiral expansion, other types 
of energy and force distributions (light-cone, IMF, etc.) can  easily be obtained from the
Breit frame ones using the  method of Abel transformations (and its generalisation for spin-1
particles \cite{MVPinprep})  \cite{Panteleeva:2021iip,Kim:2021jjf}.

Various components of the static EMT for  hadrons with arbitrary spin can be decompozed into
multipoles of the hadron's spin operator. The expansion to the quadrupole
order has the following form \cite{Polyakov:2018rew,Polyakov:2019lbq,Sun:2020wfo,Panteleeva:2020ejw}\footnote{In what follows, we shall suppress the hadron's spin indices $\sigma, \sigma'$ when
their position is obvious.
Also, we employ here the parametrization of the static stress tensor that differs from that of
Refs.~\cite{Polyakov:2019lbq,Sun:2020wfo} by a simple redefinition. 
The corresponding relations are given in the appendix of Ref.~\cite{Panteleeva:2020ejw}. }:
\begin{eqnarray}
\label{eq:quadrupole00}
T^{00}({\bf r})& =&\varepsilon_0(r) +\varepsilon_2(r)\hat Q^{pq}Y_{2}^{pq}+\ldots,\\
\nonumber
T^{ik}({\bf r})& =&p_0(r) \delta^{ik}+s_0(r)Y_2^{ik} + \left(p_2(r)+\frac13 p_3(r)-\frac19 s_3(r)\right) \hat{Q}^{ik} \\
\label{eq:quadrupoleik}
&+& \left( s_2(r)-\frac12 p_3(r)+\frac16 s_3(r) \right) 
2 \left[\hat{Q}^{ip}Y_{2}^{pk}+\hat Q^{kp}Y_{2}^{pi} -\delta^{ik} \hat Q^{pq}Y_{2}^{pq}    \right]\\
\nonumber
&+& \hat Q^{pq}Y_{2}^{pq}\left[\left(\frac23 p_3(r)+\frac19 s_3(r)\right)\delta^{ik}+\left(\frac12 p_3(r)+\frac56 s_3(r)\right) Y_2^{ik}\right]  +\ldots .
\end{eqnarray}
Here, the ellipses denote the contributions of $2^n$th multipoles with $n>2$. They are absent for
a spin-1 particle.  The quadrupole operator is a $(2J+1)\times (2J+1)$
matrix:
\be
\hat Q^{ik}=\frac 12 \left( \hat J^{i} \hat J^{k}+\hat J^{k} \hat J^{i} -\frac 23 J(J+1)
\delta^{ik} \right),
\ee
which is expressed in terms of the spin operator $\hat J^{\, i}$. The spin operator can be expressed
in terms of SU(2) Clebsch-Gordan coefficients (in the spherical basis):
\be
\hat J^{\ \mu}_{\sigma'\sigma}= \sqrt{J(J+1)}\  C_{J \sigma 1\mu}^{J \sigma'}.
\ee
Furthermore, we introduce the irreducible (symmetric and traceless)
tensor  of  rank $n$ made out of ${\bf r}$:
\be
Y_{n}^{i_1 i_2 ... i_n} = \frac{(-1)^n}{(2 n-1)!!} r^{n+1} \partial^{i_1}...\partial^{i_n} \frac{1}{r}, 
\ee
i.e.
\be
Y_0=1,\quad  Y_1^{i}=\frac{r^{i}}{r}, \quad Y_2^{ik}=\frac{r^{i} r^{k}}{r^2}-\frac13 \delta^{ik}, \quad {\rm etc.}
\ee 
Note that only the monopole quantities $\varepsilon_0(r)$, $p_0(r)$, and $s_0(r)$ are left after
spin averaging. The functions $\varepsilon_0(r)$ and $\varepsilon_2(r)$ correspond to the spin-averaged
energy density and  to the quadrupole deformation of the energy density inside the hadron,
respectively. There is an obvious relation $\int d^3 r\ \varepsilon_0(r)=m $. Also, it is obvious
that $\varepsilon_2(r)=0$ for hadrons with spin 0 and 1/2 (that is why such hadrons can be
called spherically symmetric).

From the equilibrium condition for the stress tensor, $\partial_kT^{ik}({\bf r})=0$, one
can easily obtain the equations for the functions $p_n(r)$ and $s_n(r)$:
\be
\label{eq:SFE}
\frac{d}{dr} \left(p_n(r)+\frac 23 s_n(r)\right)+\frac 2r s_n(r)=0, \ \ {\rm for }\ n=0,2,3.
\ee 
To see the physical meaning of the quadrupole force distributions $p_{2,3}(r)$ and $s_{2,3}(r)$, 
it is instructive to look at the force acting on the infinitesimal 
radial area element $dS_r$  ($d \vec{ S}=dS_r\vec{e}_r+dS_\theta\vec{e}_\theta+dS_\phi\vec{e}_\phi$).
With the help of the parameterization of Eq.~(\ref{eq:quadrupoleik}) 
and the relation of the force to the stress tensor, $dF_i=T_{ik} dS_k$, we obtain:
\begin{eqnarray}
\label{Eq:force-spherical components}
	\frac{dF_r}{dS_r}&=&p_0(r) + \frac 23 s_0(r)+\hat Q^{r r}  \left(p_2(r)+ \frac 23 s_2(r)+p_3(r)+\frac 23 s_3(r)\right) , \\	
	\label{Eq:force-spherical components1} 
	\frac{dF_\theta}{dS_r}&=& \hat Q^{\theta r}  \left(p_2(r)+
                                  \frac 23 s_2(r)\right), \nonumber \\
   \frac{dF_\phi}{dS_r}&=& \hat Q^{\phi r}  \left(p_2(r)+ \frac 23 s_2(r)\right).
\end{eqnarray}
We see that in contrast to spherically symmetric hadrons, the radial area element experiences
not only normal forces but  also tangential ones. 
The strengths of the tangential forces are governed by $p_2(r)$ and $s_2(r)$, the quadrupole
force distributions $p_3(r)$ and $s_3(r)$ contribute to the spin-dependent part of the radial force. 

Using the result for the $t$-dependence of GFFs in the chiral limit of Eq.~(\ref{ChexpFFs}) obtained
in the previous section, we can easily  calculate the analytic expressions for the large distance
behaviour (in the chiral limit) of the energy and force distributions
defined in Eqs.~(\ref{eq:quadrupole00}), (\ref{eq:quadrupoleik}):
\begin{eqnarray}
\nonumber
\varepsilon_0(r)&=&\frac{g^2_{\omega\rho\pi}}{32 \pi^2 F^2}\ \frac{1}{r^6}-\frac{3(2 a+5 g^2_{\omega\rho\pi})}{32\pi^3 F^2 M_R} \ \frac{1}{r^7}
+O\left(\frac{1}{r^8}\right) \,, \\
\nonumber
\varepsilon_2(r)&=&\frac{3 g^2_{\omega\rho\pi}}{128 \pi^2 F^2}\ \frac{1}{r^6}+\frac{21(4 a-5 g^2_{\omega\rho\pi})}{256\pi^3 F^2 M_R} \ \frac{1}{r^7}
+O\left(\frac{1}{r^8}\right) \,, \\
\nonumber
p_0(r)&=&-\frac{g^2_{\omega\rho\pi}}{96 \pi^2 F^2}\ \frac{1}{r^6}+\frac{(16 a+15 g^2_{\omega\rho\pi})}{144\pi^3 F^2 M_R} \ \frac{1}{r^7}
+O\left(\frac{1}{r^8}\right) \,, \\
\label{eq:pslarger}
s_0(r)&=&\frac{g^2_{\omega\rho\pi}}{32 \pi^2 F^2}\ \frac{1}{r^6}-\frac{7(16 a+15 g^2_{\omega\rho\pi})}{384\pi^3 F^2 M_R} \ \frac{1}{r^7}
+O\left(\frac{1}{r^8}\right) \,, \\
\nonumber
p_2(r)&=&\frac{g^2_{\omega\rho\pi}}{32 \pi^2 F^2}\ \frac{1}{r^6}{  +\frac{5(20 a-13 g^2_{\omega\rho\pi})}{192\pi^3 F^2 M_R} \ \frac{1}{r^7} }
+O\left(\frac{1}{r^8}\right) \,,\\
\nonumber
s_2(r)&=&-\frac{3g^2_{\omega\rho\pi}}{32 \pi^2 F^2}\ \frac{1}{r^6} {  - \frac{35(20 a-13 g^2_{\omega\rho\pi})}{512\pi^3 F^2 M_R} \ \frac{1}{r^7} }
+O\left(\frac{1}{r^8}\right) \,, \\
\nonumber
p_3(r)&=&-\frac{9g^2_{\omega\rho\pi}}{128 \pi^2 F^2}\ \frac{1}{r^6}{ -} \frac{7(8 a-5 g^2_{\omega\rho\pi})}{48\pi^3 F^2 M_R} \ \frac{1}{r^7}
+O\left(\frac{1}{r^8}\right) \,, \\
\nonumber
s_3(r)&=&\frac{27g^2_{\omega\rho\pi}}{128 \pi^2 F^2}\ \frac{1}{r^6} { + } \frac{49(8 a-5 g^2_{\omega\rho\pi})}{128\pi^3 F^2 M_R} \ \frac{1}{r^7}
+O\left(\frac{1}{r^8}\right)\,.
\end{eqnarray}
In Refs.~\cite{Polyakov:2018zvc,Perevalova:2016dln,Lorce:2018egm} it was conjectured that for the
stability of a mechanical system  the spin averaged pressure and shear forces should 
satisfy the inequality
\be
\label{eq:stab1}
\frac{2}{3} \, s_0(r) +p_0(r) \geq 0\,,
\ee
which corresponds to positivity of the radial pressure.
From the derived large distance behaviour of the $p_0(r)$ and $s_0(r)$, we see that the
inequality Eq.~(\ref{eq:stab1}) is indeed satisfied. However, the $\rho$-meson
decays in our theory. The terms in the large distance expansion (\ref{eq:pslarger}) which
``know'' about the instability of the particle are proportional
to the $\rho\pi\pi$ coupling constant squared $\sim a$. It is interesting to note that the corresponding
terms violate the stability condition of Eq.~(\ref{eq:stab1}), also
the corresponding terms in the spin averaged energy density $\varepsilon_0(r)$ violate its positivity. 
A detailed study
of the relations between mechanical stability conditions and the decay of unstable
particles will be given elsewhere.

\section{Summary}
\label{summary}
To summarize, we have applied chiral EFT to vector
mesons and Goldstone bosons in the presence of an external gravitational field. 
Using standard definitions,  we obtained the expressions of the EMT in flat background metric.
As first noticed in Ref.~\cite{Donoghue:1991qv}, terms in the effective Lagrangian involving
the gravitational curvature, which vanishes 
in the flat background, do give non-trivial contributions to the EMT.
This also happens for the case at hand. Therefore, in order to keep
track of all relevant contributions to the EMT in flat spacetime, it is necessary
to consider effective Lagrangian in curved spacetime. 
In the  next step,  we calculated the gravitational form factors of the $\rho$-meson at
next-to-leading order. This involves the calculation of tree-level and one-loop diagrams. 
To get rid of ultraviolet divergences and power counting violating pieces we applied the complex-mass
renormalization scheme, which allows one to subtract also the large imaginary parts from loop diagrams. 
The matrix element of the EMT for this spin-1 hadrons is parameterized using six independent structures.    
We do not give the rather lengthy expressions of the obtained expressions of the gravitational form
factors of the $\rho$-meson  in terms of standard loop
functions\footnote{These are available from the authors upon
  request.}, but focus on the chiral 
expansion of the form factors at zero momentum transfer and of their slopes. These expressions
should be useful for lattice extrapolations of the corresponding results by taking the pion mass
to its physical value. 
We also presented the expansion of the form factors in the small-$t$ region in the chiral limit. 
Using these expression, we further calculated the large-distance behaviour of the 
energy distribution and the internal forces. 
The obtained results are consistent with the  stability condition of a mechanical system.

\section*{Acknowledgments}
The authors thank Julia Panteleeva for checking some equations. 
This work was supported in part by BMBF (Grant No. 05P18PCFP1),  by Georgian Shota Rustaveli National
Science Foundation (Grant No. FR17-354),   
by the NSFC and the Deutsche Forschungsgemeinschaft through the funds provided to the Sino-German
Collaborative Center TRR110 ``Symmetries and the Emergence of Structure in QCD"
(NSFC Grant No. 12070131001, DFG Project-ID 196253076-TRR 110)
and by the EU Horizon 2020 research and
innovation programme, STRONG-2020 project under grant agreement No. 824093. 
The work of UGM was also supported in part by the Chinese Academy of Sciences
(CAS) through a President's International Fellowship Initiative (PIFI) (Grant No. 2018DM0034),
by the VolkswagenStiftung (Grant No. 93562).

\end{document}